\documentclass{article}

\usepackage{arxiv}

\usepackage[utf8]{inputenc} % allow utf-8 input
\usepackage[T1]{fontenc}    % use 8-bit T1 fonts
\usepackage{hyperref}       % hyperlinks
\usepackage{url}            % simple URL typesetting
\usepackage{booktabs}       % professional-quality tables
\usepackage{amsfonts}       % blackboard math symbols
\usepackage{nicefrac}       % compact symbols for 1/2, etc.
\usepackage{microtype}      % microtypography
\usepackage{lipsum}		% Can be removed after putting your text content
\usepackage{graphicx}
\usepackage{natbib}
\usepackage{doi}

\usepackage{orcidlink}
\usepackage[affil-it]{authblk}
\usepackage{placeins}

\title{Decision-Making Amid Information-Based Threats in Sociotechnical Systems: A Review}

\author[1]{Aaron R. Allred\,\orcidlink{0000-0001-5241-2830}}
\author[1]{Erin E. Richardson\,\orcidlink{0009-0005-3839-4759}}
\author[2]{Sarah R. Bostrom\,\orcidlink{0000-0001-6527-5978}}
\author[3]{James Crum\,\orcidlink{0000-0001-7806-8572}}
\author[3]{Cara Spencer\,\orcidlink{0000-0002-8743-6960}}
\author[4]{Chad Tossell\,\orcidlink{0000-0003-1662-9308}}
\author[2]{Richard E. Niemeyer\,\orcidlink{0000-0002-1058-253X}}
\author[3,5]{Leanne Hirshfield\,\orcidlink{0000-0003-0111-6948}}
\author[1]{Allison P.A. Hayman\,\orcidlink{0000-0001-7808-8557}}

\affil[1]{Smead Department of Aerospace Engineering Sciences, University of Colorado, Boulder, CO, USA}
\affil[2]{United States Air Force Academy, USAFA, CO, USA}
\affil[3]{Institute of Cognitive Science, University of Colorado, Boulder, CO, USA}
\affil[4]{Department of Human Factors, Safety \& Social Sciences, Embry-Riddle Aeronautical University, Daytona Beach, FL, USA}
\affil[5]{Department of Computer Science, University of Colorado, Boulder, CO, USA}

\makeatletter
\renewcommand{\maketitle}{%
  \begin{center}
    {\LARGE \bfseries \@title \par}
    \vskip 1em
    {\large \@author \par}
    \vskip 1em
    {\@date \par}  % <-- added this line to show the date
    \vskip 2em
  \end{center}
}
\makeatother

\renewcommand{\shorttitle}{Decision-Making Amid Information Threats}

\hypersetup{
pdftitle={Decision-Making Amid Information-Based Threats in Sociotechnical Systems: A Review},
pdfsubject={cs.HC, cs.CY},
pdfauthor={Aaron R.~Allred, Erin E.~Richardson, Allison A.P.~Hayman},
pdfkeywords={Human-AI Teaming, Decision Making, Cognition, Cognitive Security},
}

\begin{document}
\maketitle

\begin{abstract}
	 Technological systems increasingly mediate human information exchange, spanning interactions among humans as well as between humans and artificial agents. The unprecedented scale and reliance on information disseminated through these systems substantially expand the scope of information-based influence that can both enable and undermine sound decision-making. Consequently, understanding and protecting decision-making today faces growing challenges, as individuals and organizations must navigate evolving opportunities and information-based threats across varied domains and information environments. While these risks are widely recognized, research remains fragmented: work evaluating information-based threat phenomena has progressed largely in isolation from foundational studies of human information processing. In this review, we synthesize insights from both domains to identify shared cognitive mechanisms that mediate vulnerability to information-based threats and shape behavioral outcomes. Finally, we outline directions for future research aimed at integrating these perspectives, emphasizing the importance of such integration for mitigating human vulnerabilities and aligning human-machine representations. 
\end{abstract}

% keywords can be removed

% \newpage
% \tableofcontents  % <-- This generates the ToC automatically
% \bigskip
% \newpage

\section{Introduction}
The pervasive integration of digital information systems into human information sourcing and communication has transformed how information influences judgment and behavior. The unprecedented scale and reach of these systems have expanded both the opportunities for beneficial information exchange and the exposure to information-based threats, such as the dissemination of erroneous or misleading content. These information-based threats may pose substantial risks to societal, political, and economic stability, prompting over a decade of research into their effects \citep{muhammed_t_disaster_2022, broda_misinformation_2024}. Particularly concerning are the potential impacts on human cognitive processes, which can undermine individual and organizational functions across diverse tasks and operational settings \citep{adams_why_2023}. Despite this recognition, debate persists regarding whether such threats are effective at scale \citep{budak_misunderstanding_2024}, especially given that they may constitute a small fraction of consumed information \citep{allen_evaluating_2020}. However, the information landscape continues to transform, with increased human reliance on large language models, which often produce erroneous yet convincing responses \citep{kim_fostering_2025}. Yet, there remains an insufficient focus on understanding how these threats may influence behavior \citep{allen_addressing_2025}. 

Compounding this challenge, relevant bodies of work have evolved in isolation from one another. Over fifty years of literature, beginning with \citet{tversky_judgment_1974}’s seminal work on cognitive heuristics and culminating in contemporary applied psychology fields (such as engineering psychology and human factors engineering), has not been integrated with recent research evaluating information-based threats at the intersection of the social and cognitive sciences. Contributing to this gap, much of the literature examining information-based threats has focused narrowly on using veracity discernment as an outcome, which queries a proxy of underlying truth judgments to assess individuals’ ability to judge the truthfulness of information. Due to a difficulty in assessing the influence of these effects, recent work in this domain has called for a shift towards measuring behavioral outcomes \citep{allen_addressing_2025}. Furthermore, studies evaluating veracity discernment lack real-world context and consequence, implicating a need for ecological validity in empirical studies \citep{crum_misinformation_2024}. In contrast, extensive literature on human information processing and broader decision-making (reviewed in \citep{wickens_engineering_2015}), including human factors engineering, a promising discipline for tackling these threats \citep{endsley_combating_2018,karwowski_grand_2025}, has scarcely examined the unique effects of information-based threats. While these fragmented efforts each bring valuable strengths, their separation has precluded the development of a cohesive human-level understanding of the effects of information-based threats on human behavior.  

Understanding the cognitive mechanisms through which information-based threats affect human behavior is crucial for two primary reasons. First, as humans increasingly rely on artificially sourced and generated information, exemplified by the large-scale adoption of large language models as assistants \citep{xi_rise_2023}, protecting human decision-making from undue influence becomes essential for individual and organizational effectiveness. Second, and more broadly, achieving representational alignment between human and artificial cognition \citep{sucholutsky_getting_2024}, a goal necessary for building AI systems that reflect human values and support human information processing and decision-making, requires a unified understanding of how humans actually process information and make decisions across contexts. Without integrating our fragmented knowledge, we cannot build technologies that align with dynamic human mental models and values, nor monitor and identify potential vulnerabilities and offer support. Insights can in turn inform the design of more robust and resilient sociotechnical systems, encompassing human teams, human-AI teams, and human-in-the-loop systems.

\begin{figure}[htbp]
    \centering
    \includegraphics[scale = 0.85]{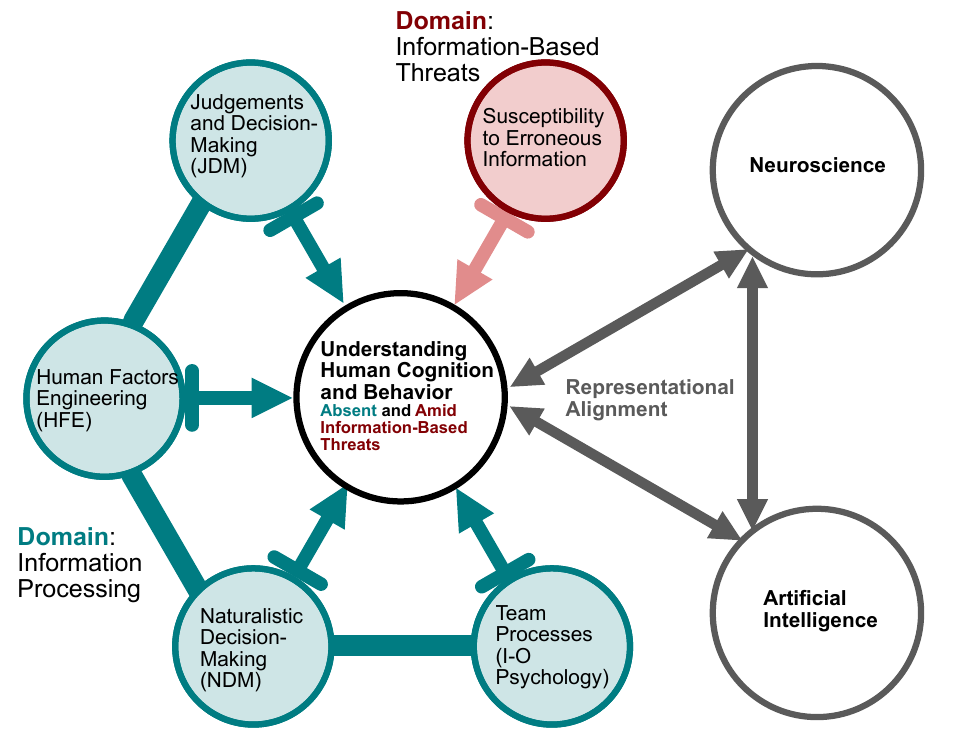}
    \caption{An overview of this paper's literature review, synthesizing fields across human psychology to better understand decision-making in sociotechnical systems in modern contexts. In the information processing domain (connected by solid, teal lines), much empirical and theoretical work has been conducted, spanning numerous sub-fields (indicated by the smaller circles: i.e., JDM, HFE, NDM, and I-O Psychology). In the information-based threats domain (red coloring), a primary focus over the last decade, a new field has emerged with complementary findings and theories. The integration of these domains (the central circle and focus of this review) provides an improved understanding of human cognition and behavior broadly. Where this effort fits in with the larger goal of aligning (and securing) artificial intelligence systems and neuroscience with human cognitive mechanisms and behavior is also depicted (three white circles, which all influence each other; represented with bidirectional arrows; reviewed in \citep{sucholutsky_getting_2024}).}
    \label{fig:Figure1}
\end{figure}

Here, we synthesize diverse research on human factors-based decision-making with recent research examining information-based threats, with primary emphasis on individual-level mechanisms while also examining team-level considerations (Figure \ref{fig:Figure1}). Both the information-based threats and the human information processing literature point to the same underlying cognitive mechanisms. This review leverages that convergence to provide an improved understanding of human decision-making across sociotechnical systems, amid and absent information-based threats. Furthermore, this approach can reveal gaps between these domains that warrant further investigation. By identifying shared mechanisms and critical gaps, we establish requirements for integrative frameworks that can capture how these mechanisms (often categorized as social, cognitive, and affective factors) shape vulnerability to information-based threats. Such frameworks are essential for developing means of maintaining sound judgments and decision-making, broadly benefiting sociotechnical systems in an increasingly complex information world. Finally, we frame this integrative pursuit as necessary for achieving representational alignment between human and artificial cognition, necessary for building technologies that safeguard human decision-making.

\section{Empirical Findings Across Domains}

\subsection{Information-Based Threats and Cognitive Security}
Information-based threats refer to information presentations that have the potential to negatively influence cognition and behavior. These threats take multiple forms, including but not limited to the direct presentation of erroneous information, a primary focus of current social psychology literature. Beyond presenting erroneous content, information can influence human information processing indirectly, such as by strategically altering perceived values associated with key outcomes (detailed in the following sections). Subsets of information-based threats are often labeled misinformation (erroneous information regardless of intent \citep{roozenbeek_psychology_2024}), disinformation (erroneous information spread intentionally \citep{lewandowsky_misinformation_2013}), and malinformation (truthful information weaponized for harm \citep{baines_ning_2020}). Among these, malinformation remains an understudied subset of information-based threat presentations in the current literature.

To defend against information-based threats, a recent field has emerged, loosely referred to as \textit{cognitive security}. In its most general form, cognitive security efforts are often self-described as the protection of \textit{human cognitive processes} from manipulation and exploitation, most commonly defined and used within a military context.  West Point’s recently developed Cognitive Security Research Lab currently defines cognitive security as maintaining rational decision-making under adversarial conditions \citep{army_cyber_institute_army_2024}. In another nascent program, NATO’s Allied Command Transformation defines a similar concept, cognitive resilience, as including protecting military capabilities and decision-making from manipulation \citep{noauthor_nato_2024}. Other researchers have also described cognitive security as involving the capability to detect, recognize, control, and counter information aimed at influencing individuals \citep{grahn_cognitive_2024} or the ability to counter information-based harms \citep{janzen_cognitive_2022}. Outside of protecting human cognition directly, cognitive security has been defined as the application of cognitive processes to cybersecurity methods \citep{andrade_cognitive_2019}, contrasting with applications primarily aiming to protect aspects of human cognition. Similarly, \citet{casino_unveiling_2025} has placed the emphasis of cognitive security on self-aware and self-learning artificial intelligence systems. More recently, researchers have proposed a domain- and species-agnostic field-level conceptual definition, describing cognitive security as “the state of having trusted boundaries protecting cognitive assets against all forms of unauthorized influence or access” \citep{ask_cognitive_2025}. This broad characterization highlights the applicability of cognitive security principles to both human and non-human systems. 

Despite these important conceptual advances, the field of cognitive security today lacks middle-level constructs. Conceptual field-level definitions can be too minimal to tie to empirical phenomena \citep{gerring_putting_2003}, offering little connection to measurable behavioral outcomes or attributes that would allow one to assess how secure the sociotechnical system truly is.  This conceptual ambiguity hinders empirical comparison across research fields (e.g., social science, cognitive science, and human factors), including the cumulative integration of findings. Recent critiques of psychological theorizing stress that many theories lack rigorous constraints, show weak linkage between constructs and measures, and often fail to specify the functional relationships among observable processes and outcomes \citep{eronen_theory_2021, bringmann_back_2022}. Consequently, understanding the cognitive mechanisms through which information-based threats shape human behavior remains essential for advancing cognitive security as a theoretical construct that can be operationalized for measurement. In reviewing evidence on the cognitive mechanisms by which information-based threats shape human behavior, linking threat-focused research with broader insights from human information processing literature, this effort may serve to advance the human psychological dimension (and associated constructs) within the field of cognitive security.

\subsection{The Influence of Information-Based Threats on Cognition and Behavior}
To understand the influence of information-based attacks, we began by reviewing research examining how human cognition (i.e., latent information processes) and behavior (i.e., their observable outcomes) are influenced by explicit presentations of erroneous information. Thus far, the influence of erroneous information at the human level has mostly focused on assessing specific belief-informed truth judgments and, to a lesser extent, individual sharing of and engagement with erroneous information \citep{murphy_what_2023}. In doing so, current research has found the effects of erroneous information to be influenced by various factors, categorized as cognitive, affective, and social (reviewed in \citep{ecker_psychological_2022}).

From these factors, humans demonstrate a number of phenomena when presented with information-based threats. For instance, people are truth-biased, demonstrating a propensity to believe incoming information \citep{levine_accuracy_1999}, which persists independently of cognitive workload \citep{pantazi_power_2018}. Truth bias has been referred to as metacognitive myopia due to a demonstrated tendency to ignore contextual information regarding a message’s veracity \citep{fiedler_chapter_2012, pantazi_power_2018}. Beyond truth bias, repetition of information (erroneous or true) has been found to influence veracity discernment performance, with repeated information perceived as truer, referred to as ‘illusory truth’ \citep{hasher_frequency_1977,udry_illusory_2024} when recognized as repeated \citep{bacon_credibility_1979}. The effect size of illusory truth relies on the amount of previous exposure, with repeated subsequent exposures in a laboratory setting yielding a diminishing but increasing effect on statements' perceived truthfulness \citep{hassan_effects_2021}. Fluency, the ease of recalling information, has been suggested to play a role in the size of the illusory truth effect \citep{reber_effects_1999,unkelbach_truth_2019} in addition to the familiarity and coherence (i.e., degree to which it is linked to references in memory) of the message \citep{arkes_generality_1989,unkelbach_truth_2019,unkelbach_referential_2017}. These effects reveal how information presentation plays a critical role in how it is processed.

Notably, several interventions have been found to be effective in reducing these effects. Concerning the illusory truth effect, warnings of potentially untrue information can result in more unbiased truth judgments \citep{jalbert_only_2023}. More broadly, skepticism and mistrust have been attributed to reducing the influence of erroneous information \citep{lewandowsky_misinformation_2012,mayer_suspicious_2011,schul_value_2008}. Additionally, increased deliberation has been found to improve the discernment of erroneous information, such as the presentation of false headlines \citep{bago_fake_2020}, further suggesting that intuitive (vs. analytic) processing of information alters vulnerability to erroneous information. Moreover, a metacognitive intervention, prompting users to evaluate their avoidance of contradictory information to their prior beliefs, has been found to be effective in improving veracity discernment performance, whereas engaging with veridical information alone does not \citep{tanaka_beyond_2025}. Despite these means to regulate and mitigate the influence of information-based threats, erroneous information in memory has been found to influence later judgments even following correction, referred to as the continued influence effect (CIE) \citep{johnson_sources_1994,lewandowsky_misinformation_2012}. 

Socially, the source of information plays a role in the extent to which it is used toward formulating judgments. For instance, higher perceived source credibility results in judgments being shaped more by erroneous information from those sources \citep{nadarevic_perceived_2020}. The perception of source credibility has been attributed to several perceived social factors. People trust sources that are perceived to share their worldviews \citep{pennycook_lazy_2019,pennycook_psychology_2021}. Similar to the effects of deliberation, inaccurate truth judgments are associated with intuitive thinking when information comes from politically discordant sources, attributed to a lack of careful reasoning \citep{pennycook_lazy_2019,pennycook_psychology_2021}. 

Beyond these factors, heightened emotionality has been found to increase vulnerability to information-based threats. Because individuals using feelings over reason to make emotionally satisfying decisions demonstrate reduced veracity discernment performance, reliance on affect has been attributed as a causal predictor of believing erroneous information \citep{martel_reliance_2020}. This finding supports how affect can influence judgments and decisions during the simultaneous presentation of erroneous information, even if the specific emotion does not directly pertain to the content assessed. Similarly, individuals attributed with more emotionality were associated with higher engagement with erroneous information \citep{horner_emotions_2021}.

Outside of truth judgments (primarily assessed through veracity discernment tasks), vulnerability to erroneous information has also been studied at the level of decision-making, particularly regarding the decision to propagate erroneous information following exposure. While correlations between discernment and behavior have been found \citep{bierwiaczonek_role_2022,stasielowicz_continuous_2022}, veracity discernment performance outcomes and sharing behavior can at times contradict one another. A recent study found that 16\% of headlines identified as false were purposefully shared anyway \citep{pennycook_psychology_2021,pennycook_shifting_2021}.  Further, older adults demonstrate a greater propensity to share false information on social media despite being better at discerning erroneous information than their younger adult counterparts, the reason for which has yet to be fully explained \citep{brashier_aging_2020,van_der_linden_using_2023}.  In another key finding, individuals made aware of their own accuracy assessments reduced erroneous information sharing \citep{pennycook_shifting_2021}, suggesting metacognition plays a role in regulating the propagation of erroneous information through improved self-attention to the accuracy of one’s judgments. Despite these prior examinations of sharing behavior, outcomes in the presence of erroneous information are largely aimed at evaluating truth judgments using veracity discernment tests \citep{murphy_what_2023}. Veracity discernment is an attractive experimental outcome in that it is easily observable and allows experimenters to answer novel hypotheses while conducting carefully controlled experiments studying human behavior \citep{pennycook_practical_2021}. However, the widespread reliance on veracity discernment as a measure has limited the field’s ability to fully capture the broader behavioral impacts of information-based threats, which may extend beyond information sharing.

\subsection{Judgments and Decision-Making Under Uncertainty}
Aside from these findings surrounding information-based threats, the last fifty years of research in fields related to judgments and decision-making (JDM) have revealed key insights into how humans process information to make judgments and decisions more generally. This body of literature has revealed a number of suboptimalities in human information processing when compared to an ideal (i.e., normative) observer, which provides an upper-bound benchmark on optimal human behavior. Even under veridical information conditions, humans often utilize intuitive thinking, relying on heuristics and demonstrating biases when formulating judgments under uncertainty \citep{tversky_judgment_1974}. Mirroring findings in the literature examining information-based threats, these deviations from normative predictions are in turn influenced by cognitive, affective, and social factors.

One of the most influential mediators of judgment is affect \citep{forgas_mood_1995,greifeneder_when_2011,lerner_emotion_2015}. For instance, judgments of perceived negative outcomes are overestimated following incidental negative affect, induced by unrelated, negatively charged content, and vice versa \citep{johnson_affect_1983}. Moreover, when utilizing judgments to make decisions, chosen courses of action also deviate from objective ideals (i.e., normative expected utility theory) even when probabilities are provided and clearly stated \citep{kahneman_prospect_1979,tversky_advances_1992}. To capture these deviations, the descriptive models of Prospect Theory (Original and Cumulative) transform actual probabilities into subjectively weighted probabilities, and perceived gains and losses are modeled with asymmetric benefit and value functions, respectively. These modeling approaches enable descriptive models of choice that are congruent with observed human behaviors. In a following landmark study, subjective risk and benefit evaluations were found to be modulated by integral affect \citep{finucane_affect_2000}. Whereas actual benefits and risks typically vary proportionately in an environment, perceived benefits and risks instead often vary inversely \citep{alhakami_psychological_1994}. Specifically, positive affect yields high-benefit and low-risk evaluations, and negative affect yields the opposite. Further, information that alters the favorability of a person’s affective evaluation systematically alters risk and benefit judgments necessary for choosing an action \citep{finucane_affect_2000}. Based on these findings, this modulatory relationship can be linked directly to choice evaluations \citep{lerner_emotion_2015,phelps_emotion_2014,schulreich_fear-induced_2020}. Moreover, outcome-associated values are goal-dependent \citep{molinaro_intrinsic_2023}. These findings highlight that affect systematically shapes goal-oriented risk and benefit evaluations, altering decision-making from normative expectations.

Socially, a number of factors have been found to influence decision quality, including information sharing, which is often necessary for formulating sound decisions with incomplete information. Within groups, decisions can be influenced by biased information sampling during discussions \citep{stasser_pooling_1985}. Further, the sources of information have been found to dictate decisions, with sources of greater perceived authority given more weight \citep{schobel_social_2016}. Messages attributed to in-group members are more persuasive than alternatives \citep{mackie_processing_1990}, and peripheral cues such as perceived attractiveness, similarity to the message receiver, and power play a role in modulating source credibility \citep{brinol_source_2009}. During tasks that require information sharing, remote deliberation amongst team members has been found to reduce decision quality compared to in-person deliberation \citep{javalagi_zooming_2024}. Collectively, this evidence demonstrates that social context can mediate judgments, highlighting the importance of interpersonal dynamics even in individual decision-making.

Crucially, these deviations are not uniformly applicable across contexts and paradigms. The extent of cognitive resources and metacognitive regulation utilized by individuals when formulating judgments and making decisions has been found to play a substantial role in regulating the degree of deviations from normative predictions. For statistically based judgments, individuals can neglect base rates when establishing priors \citep{tversky_judgment_1974} or over-rely on base rates for belief-based judgments \citep{pennycook_base_2014}. However, the degree to which the representativeness heuristic is utilized can be modulated by cognitive interventions. For example, ‘easy fix’ interventions aimed to redirect attention toward relevant cues with additional deliberation time \citep{boissin_easy-fix_2024} can align judgments toward logical formulations. Supporting this idea, the degree to which working memory is utilized has been proposed by some as the delineator of System 1 (intuitive) and System 2 (analytical) processing \citep{evans_dual-process_2013,evans_dual-processing_2008}. Building on findings that deliberate thinkers have an advantage over quick, intuitive thinkers \citep{mata_metacognitive_2013}, metacognition has been proposed as the mediator of System 1 and System 2-driven reasoning and behaviors \citep{de_neys_perspective_2017,evans_dual-process_2009}. 

\subsection{Naturalistic Decision-Making}
Many of the findings reviewed so far are derived from laboratory-based findings. Another descriptive understanding, Naturalistic Decision Making (NDM), aims to understand how people make decisions in real-world settings, which often involve uncertainty, high stakes, limited time, ambiguous objectives, and instability. Outside laboratory settings, a number of \textit{ecological} factors (compounding the previously categorized social, affective, and cognitive factors) have been found to affect the decision-making process. These factors include dynamic, uncertain environments, shifting and/or competing goals, and time pressure \citep{orasanu_reinvention_1993}. Other naturalistic factors, such as physical fatigue and physical encumbrance (leading to fatigue) encountered when donning task-specific equipment, have been found to influence decision-making \citep{levitt_multiple_1971,eddy_effects_2015}. These factors may be generally attributed to decreased available cognitive resources from the constraints of a task.

Of these factors, time pressure presents one of the most notable ecological constraints influencing judgment and decision-making. Decision-making can be characterized by the extent to which it involves intuitive versus analytic processing, and time pressure and stress can play a crucial role in determining this reliance \citep{hammond_effects_1997}. Under time pressure, proficient decision makers (with experience relevant to their task setting) rely more on pattern recognition and memory retrieval to generate courses of action, contrasting sharply with the extensive alternative evaluation processes typical of novices. For example, expert chess players often act on the first retrieved move, while novices engage in exhaustive searches for the best option \citep{, klein_characteristics_1995}. Similarly, experienced firefighters, ship commanders, infantry officers, and pilots adopt comparable decision strategies when facing time pressure, information uncertainty, or vague goals \citep{lipshitz_taking_2001}. These strategies combine intuitive pattern matching with mental simulation of potential outcomes \citep{ klein_naturalistic_2008}. Given that considering more alternatives increases decision time \citep{hick_rate_1952, hyman_stimulus_1953}, this reliance on memory-based intuition represents an efficient adaptation to high-stakes decisions under constraints by effectively using consolidated knowledge to circumvent real-time information limitations \citep{klein_naturalistic_2015}. In comparison, information-based threats pose their greatest risks in real-world settings, which may also be time-constrained, with individuals relying on pattern recognition and memory retrieval when making judgments and decisions. 

\subsection{Team Decision-Making}

Another fundamental driver of judgments and decision-making that can be expected to vary across naturalistic environments is the role of teaming \citep{mosier_judgment_2010}. While laboratory experiments and even much naturalistic research often focus their examinations on individuals, team-based research demonstrates that behavior can differ substantially when decisions are made collectively. Teams are characterized as highly interdependent, with members sharing goals and influencing other team members’ decision-making  \citep{hollenbeck_multilevel_1995}. 

Teams present unique structures upon which to examine the influence of information-based threats, and have been studied extensively by industrial and organizational (I-O) psychologists. For instance, while often presenting as more sophisticated than individuals, even unsophisticated teams arrive at decisions in shorter times than individuals acting alone \citep{zimmermann_team_2020}. However, specific benefits of teams in decision-making depend on the team composition, cohesion, hierarchy, and specific operational context \citep{reader_team_2017}. While this article primarily examines human information processing on an individual level, both individual and team-level cognitive security are likely to be crucial mediators of team performance in the presence of information-based threats. Supporting this likely crucial role, cognitive security is closely related to behavioral processes, including team information sharing, and collective “affective and cognitive” emergent states \citep{kozlowski_work_2003,grossman_teamwork_2017} that mediate team performance (discussed in greater detail in Section \ref{sec:team_models}).

Team information-sharing, which may include the propagation of information-based threats among team members, has been noted as an important mediating behavioral process for teams. Described as a process in "which team members collectively utilize their available informational resources" \citep{mesmer-magnus_information_2009}, team information sharing is influenced by task conflict such that information sharing between members is higher during debate-framed discussion rather than disagreement-framed discussion \citep{tsai_pursuit_2016}. In innocuous informational contexts, absent information-based threats, increased team information-sharing has been found to positively relate to team performance \citep{hulsheger_team-level_2009}. In contrast, team information sharing's effect on team performance amid information-based threats likely depends on the quality of the information shared (reflecting an interaction of individual and team veracity discernment) rather than the propensity for members to share.

\subsection{Convergence Across Domains}
The empirical findings reviewed above indicate an important convergence: outcomes associated with information-based threats are shaped by the same cognitive (intuitive vs. analytic processing, metacognition), affective (emotion-modulated valuations), and social (source credibility, in-group effects) mechanisms that influence judgment and decision-making more broadly across contexts. This overlap suggests that information-based threats exploit fundamental features of human cognition rather than inducing novel phenomena. Next, we examine how existing frameworks have attempted to capture and describe these processes, revealing gaps that integration could address.

\section{Existing Frameworks and Models}

Decision-making under information-based threats poses substantial challenges for both individuals and teams. Mirroring how empirical research has documented systematic deviations from normative decision-making in these contexts, existing frameworks and models have also developed largely independently from one another. At the individual level, models focus either on general information processing or on specific threat responses. At the team level, research has examined team effectiveness broadly but has not yet addressed how teams maintain cognitive security when facing information-based threats. Here, we review notable framings at the human level, across both individuals and teams, and examine important models across both domains.

\subsection{Information-Based Threat Models}
\label{sec:threat_models}
To explain empirical data pertaining to altered veracity discernment in the presence of erroneous information, several models have used Bayesian inference to describe elements of erroneous information’s influence on cognition. Bayesian network models have been used to quantitatively explain belief polarization by modeling how worldview can influence priors and evidence evaluations \citep{cook_rational_2016}, leading to updated posteriors. Building on this work, but using a conceptual framework, a Bayesian inference paradigm has been constructed to depict individual receptivity to erroneous information, suggesting that altered evidence evaluations (i.e., a likelihood evaluation) depend on perceived source credibility and information plausibility \citep{zmigrod_misinformation_2023}. However, these models focus solely on examining updates to beliefs and do not model the role of affect and other descriptive decision-making elements, particularly towards task-relevant objectives. 

Similarly, a three-part framework by \citet{brashier_judging_2020} conceptually describes the illusory truth effect (a phenomenon associated with misinformation processing) through a “Bayesian-like” base-rate alteration of priors, influence of feelings, and consistency with information retrieved from memories. However, because their framework does not capture relationships between evidence evaluation for posterior updates and decision-making over time, it does not necessarily generalize to additional information-based threat phenomena or enable quantitative comparisons to temporal predictions of illusory truth responses.

Addressing broader aspects of information processing amid information-based threats, \citet{amazeen_misinformation_2024} provides a framework that examines both the antecedents and consequences of erroneous information recognition and response. It integrates an individual's dispositional factors (such as the propensity to engage in thinking activities) and situational factors (such as mental resources and domain-specific knowledge), with information and intervention presentation formatting and an individual's cognitive coping strategies to explain how individuals recognize and respond in unique contexts. While this framework addresses the broader context of information processing in these scenarios, including representing multifaceted outcomes (categorized as cognitive, affective, and behavioral outcomes), it remains primarily descriptive and does not provide a quantitative formalization to describe how these factors interact over time to influence judgments and decision-making.

Outside of social and cognitive psychology, some work has adopted alternative cognitive framings. Evaluating information-based threats through a human factors lens, \citet{endsley_combating_2018} highlights how information attacks undermine situation awareness, distort mental models, and overload cognitive resources (such as attention) that are essential for sound decision-making. Here, she emphasizes the importance of designing systems that support accurate perception, comprehension, and projection (the core elements of situation awareness \citep{endsley_toward_1995}) and argues that countering these effects requires new cognitive engineering and human factors approaches designed to maintain well-informed human decision-making in complex information environments, such as the public policy information space. Despite \citet{endsley_combating_2018}’s contributions, little additional progress has been made in applying these approaches or in leveraging advances from the information processing domain more broadly, which we discuss in the following subsection.

\subsection{Information Processing Models}
Extending beyond capturing updates to beliefs towards task-relevant judgments and decisions, the fields of engineering psychology and human factors engineering (HFE) have leveraged the information processing model (IPM) \citep{wickens_engineering_2015,wickens_information_2021} for describing the mediating factors influencing human performance. The IPM describes how individuals filter and process incoming environmental information to select an action, which may alter the information environment for future processing and choice. Specifically, \citet{wickens_engineering_2015} outlines this flow of information as the way people attend, perceive, think, remember, decide, and act, mirroring the full information processing chain as described by strategist \citet{boyd_essence_1995}'s observe, orient, decide, and act (OODA) Loop, with origins in cybernetics, systems theory, and strategy. The IPM is useful for understanding a coarse interplay of information, cognitive structures, cognitive resources, and the formulation of judgments and decisions. However, the IPM does not explicitly model the influence of cognitive constraints leading to heuristics and biases, social factors such as perceived source credibility, nor the influence of affect. As a result, it has yet to be directly utilized to quantify the role of information-based attacks on judgments and decisions. Further, the IPM does not contain parameterized relationships, making it ill-suited for hypothesis generation and tuning with empirical data.

While the fields of engineering psychology and human factors have leaned into the IPM for its descriptive benefits, the Adaptive Control of Thought--Rational (ACT-R) model has been used by cognitive scientists as a computational model of human cognition \citep{anderson1998atomic,anderson2004integrated,anderson2007physical}. ACT-R is a hybrid cognitive architecture that models human cognition by integrating symbolic production rules with subsymbolic learning and memory mechanisms (recently reviewed in \citep{ritter_actr_2019}). The architecture is organized into modules corresponding to distinct cognitive functions, including declarative memory, procedural memory, goal management, perception (visual and aural), and motor actions (manual and vocal). Knowledge is represented symbolically as chunks and production rules (organized as if-then statements). Competing productions are selected via value-based rules. ACT-R has been applied extensively to laboratory cognitive tasks such as memory, problem solving, and language comprehension, as well as to applied fields including human-computer interaction, air-traffic control, and intelligent tutoring systems \citep{anderson2004integrated,ritter2007cognitive,lebiere1999cognitive}. 

However, the major limitation of ACT-R is its lack of inherent mechanisms for representing social and affective factors \citep{ritter_actr_2019}, as well as the effects of cognitive resource constraints, despite providing a mechanistic account of cognitive processes such as memory retrieval, attention, and procedural skill acquisition. In an attempt to overcome the disregard of affective factors, ACT-R has recently been modified to capture the continued influence effect \citep{hough_model_2024,hough_modeling_2024}. This is achieved through the addition of a core-affect mechanism \citep{larue_core-affect_2017}, which interacts with cognitive mechanisms such that declarative memory chunks are assigned additional values associated with referencing a chunk, including an arousal term. Thus, affect here is implemented as a modulator of memory retrieval, rather than a modulator of outcome valuations (such as the case with risk and benefit evaluations \citep{finucane_affect_2000,johnson_affect_1983}). These results suggest promising future avenues of integration with information-based threat frameworks and outcome-associated valuations that drive action selection (such as in Prospect Theory). 

Despite this progress, a number of limitations remain. ACT-R has been criticized in that it is overparameterized, requiring detailed tailoring to specific tasks \citep{sun_cognition_2006}. Thus, ACT-R is well-suited for modeling task-focused cognition under well-described conditions, but it struggles to capture emergent, context-sensitive, or affectively modulated behaviors, which are central mediators to judgments and decision-making under uncertainty and information-based threats. Finally, ACT-R is not well-suited for handling uncertain, probabilistic representations of information (e.g., uncertainty or ambiguous cues) that may arise on the stimulus and neural levels (reviewed in \citep{faisal_noise_2008}), likely a key component of processing information-based threats in uncertain information environments.

Another popular cognitive architecture, the "state, operator, and result" (Soar) model shares several limitations with ACT-R for understanding human behavior while offering complementary strengths \citep{laird_soar_2012,laird_analysis_2022}. Soar does not capture affective or social factors, and developing detailed task-specific models may not generalize and be highly tailored. However, Soar is well-suited for modeling strategic, goal-driven behavior, intended for designing AI agents with complex cognitive capabilities. Broadly considering a range of models and frameworks for understanding human cognition and behavior can help identify shared gaps, particularly in modeling the social, affective, and context-sensitive influences that are critical for understanding information-based threats.

\subsection{Bayesian Cognitive Modeling}
Outside the frameworks discussed in the information-based threats literature to capture erroneous information receptivity (reviewed in Section \ref{sec:threat_models}), Bayesian modeling has been broadly used to capture information processing, well-suited for describing probabilistic representations. Bayesian inference for cognitive modeling has been applied for the purpose of representing judgments \citep{chater_probabilistic_2006,kruschke_bayesian_2010,lee_bayesian_2014,lieder_anchoring_2018}, and within perceptual neuroscience, Bayesian inference has been applied to capturing many central nervous system circuits spanning decades of application \citep{battaglia_bayesian_2003,kording_bayesian_2014,kording_bayesian_2004,kording_bayesian_2004-3,yang_interaction_2012}. These efforts demonstrate the robust applicability of Bayesian inference to modeling centrally driven behaviors. However, Bayesian inference has yet to be applied to develop a general framework that captures known factors influencing judgments and decision-making for the purpose of understanding phenomena that arise when processing erroneous information towards behaviors. This gap is particularly consequential for understanding information-based threats. While Bayesian frameworks are well-suited for modeling how individuals update beliefs in uncertain environments, they have not been applied to capture the full decision-making process (from evidence evaluation through behavioral outcomes) in the presence of information-based threats.

Bayesian processing may appear contradictory to known heuristics and bias reliance for judgments and decision-making because the latter deviates from externally computed optimal solutions by experimenters \citep{kahneman_prospect_1979,tversky_judgment_1974}. Consequently, there has been much discussion within the literature concerning whether or not humans are Bayesian optimal or suboptimal \citep{bowers_bayesian_2012,gardner_optimality_2019,jones_bayesian_2011,rahnev_suboptimality_2018}. However, these two seemingly paradoxical views can be reconciled by considering that optimal solutions computed by the brain must rely on ecological, computational, and energetic constraints \citep{gardner_optimality_2019}. From this vantage, heuristics, biases, and general intuition-driven behaviors can emerge as optimal solutions to constrained objectives \citep{jones_bayesian_2011}. For instance, recent efforts in perceptual neuroscience have formulated a unifying theory to explain seemingly contradictory behavioral biases by considering limited neural resources and their efficient coding, in a Bayesian observer model \citep{hahn_unifying_2024,wei_bayesian_2015}. In cognitive decision-making, heuristics modeled via fast-and-frugal trees (effectively constrained and approximated Bayesian models \citep{martignon_naive_2003}) have been found to excel under computational constraints when compared to more exhaustive evaluations \citep{gigerenzer_heuristic_2011,martignon_categorization_2008}. Further, modern perspectives encourage reducing the emphasis on whether or not behavior is optimal and instead focus on building detailed models that broadly explain observed behaviors \citep{rahnev_suboptimality_2018} and thus further the social sciences through solution-oriented theories and methodologies \citep{watts_should_2017}. 

\subsection{Models of Naturalistic Decision-Making}

Because decision-making under ecological constraints characteristically differs from normative strategies, while likely not fundamentally different regarding underlying mechanisms, separate models have arisen to describe decision-making in naturalistic settings. \citet{klein_recognition-primed_1993}'s Recognition-Primed Decision (RPD) model, developed based on semi-structured interviews with fire ground commanders \citep{klein_rapid_1986}, emphasizes the role of experience in naturalistic contexts, which produces proficient decision makers. People can quickly match a situation to patterns they have learned through experience, enabling fast decision-making without an extensive comparison of options (i.e., comprehensive choice). In evaluating a course of action, people imagine how it will play out, enabling them to modify it or pivot until finding an action that satisfices their needs \citep{klein_naturalistic_2008,klein_rapid_1986}. Parallel tracks of NDM research have arrived at similar conclusions \citep{lipshitz_taking_2001}. Importantly, the RPD model has been informed by studies of people in field settings, including military leaders, jurors, nuclear plant operators, and anesthesiologists \citep{klein_naturalistic_2008}.

Several complementary models with NDM framings capture various degrees of our understanding of expert judgment and decision-making in alignment with the RPD model. The Recognition/Metacognition (R/M) model \citep{cohen_metarecognition_1996} extends RPD by emphasizing metacognitive monitoring, enabling decision makers to critique, verify, and adapt their intuitions when facing atypical or uncertain scenarios. The \citet{endsley_toward_1995} model of situation awareness describes a three-level information processing process as the core to effective decision making in dynamic settings: perception of environmental cues, comprehension of their significance, and projection of their future states. More recently, the Bayesian Case Model \citep{kim_bayesian_2015} provides a generative, probabilistic computational framework for case-based reasoning; it formalizes how learned prototypes from prior experience can be used to classify and interpret novel, uncertain data. Collectively, these frameworks underscore the interplay between rapid cognition, accumulated experiential knowledge, and metacognition in expert decision making.

\subsection{Team Processes and Decision-Making Models}
\label{sec:team_models}

Beyond the individual, capturing how teams process information and make decisions is critical for identifying points of vulnerability and predicting enhanced collective behavior in collaborative contexts. One established model of team processes, behavioral mechanisms that convert team inputs into outcomes, is the IMOI (Input-Mediator-Output-Input) model, which describes teams as “complex, multilevel systems that function over time, tasks, and contexts" \citep{ilgen_teams_2005}. It emerged as an improvement over the preceding I-P-O (Input-Process-Output) model by capturing a broader range of mediators that are not necessarily processes and by adding a pathway for feedback \citep{hackman_design_1978, mcgrath_influence_1962, steiner_paradigms_1986}. Inputs may include individual-level factors (such as knowledge and training), team-level factors (such as team size and structure), and task-level factors (such as risk and complexity). Mediators may include team interactions (such as leadership and communication) or emergent cognitive and affective states at the team level. Outputs may include both performance outcomes (like decision accuracy) and team outcomes (like satisfaction and cohesiveness) \citep{ilgen_teams_2005,reader_team_2017}. The IMOI model is well-supported and has been applied to study human-AI teams \citep{degen_human-ai_2024} and human-robot teams \citep{you_teaming_2017} in addition to a variety of human-human teams, including virtual teams in the United Arab Emirates \citep{khalil_investigating_2017}, mental healthcare professionals in Canada \citep{markon_profiles_2017}, school-based leadership teams in the US \citep{choi_promoting_2024}, and teams in space and analog environments \citep{kaosaar_fantastic_2022}, among others. In more recent reviews and meta-analyses, an advancement of the IMOI has been proposed that considers features that co-evolve and belong to multiple feature groups (e.g., structural and compositional features or mediating and compositional features), emphasizing that teams are dynamic and evolve over time \citep{mathieu_century_2017}. Due to its regions encapsulating team constructs, we refer to this model as the ABCDEF model of team effectiveness \citep{mathieu_embracing_2019}. While the IMOI and ABCDEF models concern team formation, effectiveness, and existence in general (with decision quality being one tangible outcome of many varied outcomes characterizing team effectiveness), they are not focused specifically on capturing and representing team decision-making processes with an information processing lens.

Often defined as “a team process that involves gathering, processing, integrating, and communicating information in support of arriving at a task‐relevant decision” \citep{cannon-bowers_shared_1993}, team decision-making is at the center of understanding the impact of information-based threats in real-world contexts. Some research efforts have developed models of team decision-making. One model centers around shared mental models, or organized knowledge shared by members, enabling coordinated operation \citep{cannon-bowers_shared_1993,orasanu_reinvention_1993}. \citet{orasanu1990shared} extended this concept to explain functioning in novel situations and emergencies, suggesting that teams develop shared situation models that include “shared understanding of the problem, goals, information cues, strategies, and member roles." Other models employ a further characterization of teams as consensus-based or hierarchical. Consensus-based teams, such as juries, may be heterogeneous and may comprise distributed knowledge, but their members have equal votes toward the group's decision \citep{hollenbeck_multilevel_1995}. Hierarchical teams, on the other hand, involve status differences between members. Initial team decision-making modeling focused on consensus-based groups. The social decision scheme (SDS) and social transition scheme (STS) models, concerned with decisions with finite alternatives, use distributions of member preferences to model choices \citep{davis_group_1973,kerr_group_1992}. \citet{hollenbeck_multilevel_1995} sought to model hierarchical teams with distributed expertise. They adapted Brehmer and Hagafors's team lens model \citep{brehmer_use_1986} into a multilevel theory of hierarchical decision making, which presents core predictors of effective team decision making. The lens model assumes that individuals' decisions involve seeking cues, assigning weights to the cues, and evaluating the set to reach decisions. The multilevel theory of hierarchical decision making purposefully excludes team processes driving core constructs, acknowledging that there are a variety of complex processes that differ between situations and should be studied in the context of particular settings and problems. We conclude that team decision-making literature currently lacks and would benefit from a unified (if coarse) mathematical formalization of how team decision quality is broadly influenced by information-based threats, considering varied environments and task demands over time.

\section{Discussion}
\label{sec:Discussion}

Taken together, this synthesis demonstrates that insights from the judgments and decision-making literature align closely with research on vulnerability to information-based threats, highlighting their mutual relevance. Congruent with this idea, the sharing of erroneous information has been suggested to be mediated by both the affect and availability of heuristics \citep{lu_heuristic_2024}. Similarly, research has shown that intuitive thinking tendencies increase susceptibility to erroneous information \citep{ecker_psychological_2022,pennycook_psychology_2021,pennycook_shifting_2021}. Despite these insights, there remains a gap in understanding how the cognitive mechanisms that give rise to heuristics and biases explicitly shape behavior in the context of erroneous information and information-based threats more broadly. Moreover, our synthesis reveals that a shared understanding on how to represent human information processing (and its influential modulators) remains fragmented. While many context-specific frameworks and models of individual and team cognition and decision-making exist across many subdomains of psychology, they remain limited in their applicability to new contexts. As a result, comprehensive frameworks integrating information threat processing with descriptive models of decision-making, which may provide new insight into their effects on underlying cognition and observable behaviors, remain absent. 

Another key limitation of existing work is the confinement of information-based threat studies to laboratory-based environments \citep{crum_misinformation_2024}. This approach may contribute to the current discourse on the effects of information-based threats, studied without ecological validity, given that ecological factors play a crucial role in decision-making. For instance, NDM research informs us on how factors such as time pressure and vague goals result in RPD-style individual decision-making strategies in naturalistic environments with perceived high stakes. However, the interaction between NDM strategies and information-based threats remains unexplored. Do expertise-driven heuristics provide protection against misleading information, or do they create unique vulnerabilities when patterns are subtly manipulated? Future research should examine how the shift from analytic to intuitive processing under naturalistic conditions modulates susceptibility to information-based threats. Due to these alterations in decision-making strategies, future work should examine the interaction of NDM strategy and the presentation of information-based threats, which may pose unique protections and vulnerabilities. 

Beyond individuals acting in naturalistic contexts, teaming introduces another layer of complexity that likely affects how information-based threats shape real-world behaviors. While much information-based threat literature to date focuses on the influence of erroneous information presented to individuals \citep{murphy_what_2023}, and individuals interacting on social media \citep{butler_misinformation_2024, grinberg_fake_2019}, teams too increasingly interact and share information through technological media. Even the most extreme, seemingly disconnected teams (such as those in isolated, confined, and extreme environments) source and share key information for making decisions through technological devices \citep{richardson_identifying_2025}. Thus, crucial gaps remain in understanding how technology, in conjunction with team mediators and team compositional factors, shapes how teams are influenced by information-based threats. In particular, the way in which elements of team dynamics (such as those associated with the IMOI and ABCDEF models) modulate the impact of information-based threats on team effectiveness has yet to be systematically studied in empirical research.  Promisingly, team decision-making literature reveals unique mediators, referred to as team processes and emergent states \citep{mathieu_team_2008} and also influenced by the environment \citep{kaosaar_fantastic_2022}, that play a role in determining shared team action policies \citep{marks_temporally_2001}. These mediators could be critical to understanding how information-based threats are effective and can be countered in various task settings over time.  

To advance beyond these limitations, new theories and frameworks describing the influence of shared social, cognitive, and affective factors across environmental contexts are needed. Central to these efforts, frameworks must capture how information-based threats influence task-relevant behaviors to understand real-world impacts. These gaps highlight the urgent need for an integrative framework that: (1) captures shared cognitive mechanisms across information processing and threat response, (2) parameterizes relationships between mechanisms and outcomes to enable hypothesis generation and empirical testing, and (3) can be extended from the individual-level to broadly encompass diverse sociotechnical systems, including team-level (such as human-human and human-AI team) vulnerabilities. Such a framework must bridge descriptive models from the information processing domain with emerging information-based threat (and cognitive security) research while accounting for the ecological validity concerns we have outlined. Beyond understanding mechanisms towards vulnerability, this emphasis will better enable the curation of technologies (such as constructive occupational co-pilots, or partners in thought \citep{collins_building_2024}) that help protect human vulnerabilities to information-based threats. Developing this integration is essential for both protecting human decision-making in adversarial information environments and advancing human-AI alignment.

These implications extend beyond understanding decision-making amid information-based threats (depicted in Figure \ref{fig:Figure1}). To improve human-computer interaction broadly, recent large-scale efforts across cognitive scientists, neuroscientists, and researchers in machine learning have outlined the importance and challenges of improving representational alignment between humans and machines \citep{sucholutsky_getting_2024}. Central to this idea is the goal of building machines that reflect human behaviors and values \citep{amodei_concrete_2016}, and ideally human cognition \citep{collins_building_2024}. Researchers at this intersection today often employ techniques such as inverse reinforcement learning \citep{jara-ettinger_theory_2019} and reinforcement learning from human feedback to infer knowledge about expert human reward functions given features in the world. 
%Thus, the goal of these efforts is to better align human and robot behaviors within a reinforcement learning framework \citep{bellman_theory_1954}. But given the static nature of rewards within reinforcement learning, representation alignment between humans and machines is currently bound by the disregard of the social, affective, and cognitive factors that modulate human information processing over time and across contexts. 
But given the static nature of these mathematical models, representation alignment between humans and machines is currently bound by the disregard of the social, affective, and cognitive factors that modulate human information processing over time and across contexts. To advance these broader disciplines, a shared, integrative understanding of human cognition, producing observable behaviors, is needed. Without first uniting the fragmented field examining human cognition and behavior, how can the higher-level field of representational alignment -- across cognitive scientists (and psychologists more broadly), neuroscientists, and machine learning researchers -- advance? 

 A paramount goal for future human-computer interaction researchers, among others, is the development of an integrative representational framework that incorporates findings and models across the subdomains of human psychology reviewed here. Such a framework, developed at individual and team levels of analysis, may reveal new dimensions unto which dynamic human mental models and values, rather than static state transitions and reward functions, may be cast. We anticipate that such future developments will better enable future systems, including AI teammates, to adapt to their dynamic human counterparts, both amid and absent information-based threats.

\section{Acknowledgments}
The authors acknowledge the Air Force Office of Scientific Research (AFOSR), award no. FA9550-23-1-0453, for support of this research. The views expressed in this report are those of the authors and do not reflect the official policy or position of the US Air Force, Department of Defense, or the U.S. Government.

\section{Author Contributions}
A.R.A. conceived the study. 
A.R.A., A.P.A.H., and E.E.R. interpreted the literature and prepared the initial manuscript. 
S.R.B, R.E.N, and J.C. provided critical ideas and feedback that shaped the research synthesis. 
L.H, A.P.A.H., R.E.N, and C.T. led funding acquisition and project supervision. 
All authors contributed to reviewing and editing the manuscript. 
All authors approved the final version for publication.

\bibliographystyle{unsrtnat}
\bibliography{biblio}  %%% Uncomment this line and comment out the ``thebibliography'' section below to use the external .bib file (using bibtex) .





% \end{thebibliography}

\end{document}